# Sensitivity of excitonic transitions to temperature in monolayers of TMD alloys

K. Ciesiołkiewicz-Klepek[1, a)], J. Kopaczek[1], J. Serafińczuk[2], R. Kudrawiec[1, b)]

*[1]Department of Semiconductor Materials Engineering,
Wroclaw University of Science and Technology,
Wybrzeże Wyspiańskiego 27, 50-370 Wrocław, Poland*

*[2]Department of Nanometrology,
Wroclaw University of Science and Technology,
Wybrzeże Wyspiańskiego 27, 50-370 Wrocław, Poland*

[a)] e-mail address: karolina.ciesiolkiewicz@pwr.edu.pl
[b)] e-mail address: robert.kudrawiec@pwr.edu.pl

Two-dimensional transition metal dichalcogenides (TMDs) offer tunable optical and electronic properties, making them highly promising for next-generation optoelectronic devices. One effective approach to engineering these properties is through alloying, which enables continuous control over the bandgap energy and excitonic transitions. In this study, we perform temperature-dependent transmission spectroscopy on monolayers of $Mo_{1-x}W_xS_2$ and $Mo(S_{1-x}Se_x)_2$ alloys, transferred onto the core of an optical fiber and measured within a cryostat over a temperature range of 20–320 K. The use of an all-fiber configuration allowed us to probe interband transitions A, B, and C with high stability and precision. We observe a systematic redshift of excitonic transitions with increasing Se content, and a blueshift when Mo is replaced with W. These spectral shifts correlate with alloy composition and enable the tuning of bandgap energies between ~1.6 eV and ~2.0 eV at room temperature. Furthermore, we analyze the temperature sensitivity of the excitonic transitions, revealing that Se incorporation enhances thermal response in a non-monotonic manner, while W substitution results in a more monotonic and stronger temperature dependence. The splitting between A and B transitions, associated with spin-orbit coupling, also varies with composition. Our findings underscore the potential of compositional engineering in 2D TMD alloys to achieve both spectral and thermal control of optical properties, relevant for the design of robust and tunable optoelectronic systems.

**Introduction**

Two-dimensional (2D) transition metal dichalcogenides (TMDs) materials have attracted considerable attention due to their unique electronic[1–3] and optical properties[4], as well as their potential for applications in modern optoelectronic devices.[5] A key aspect in designing such devices – including light-emitting diodes, lasers, sensors, and modulators – is the ability to control the bandgap.[6,7] This control can be achieved either externally, through parameters such as temperature,[8] mechanical strain,[9] or internally, by modifying the material itself. Such internal modifications include the introduction of defects[10–12] or dopants,[13,14] variation in the number of layers[15,16], or alloying.[13,17,18] For instance, it is well known that reducing the thickness of TMDs from bulk to a monolayer changes their band structure from indirect to direct.[19,20] Commonly studied TMDs such as $MoS_2$ and $WS_2$ exhibit a fundamental direct bandgap at the K point of the Brillouin zone in the monolayer limit, with pronounced optical transitions labeled A and B.[21,22] These transitions arise from the valence band splitting caused by strong spin-orbit coupling.[23]

One promising route for tailoring the properties of 2D materials is through alloying, either on the cationic or the anionic sublattice.[13] This allows for continuous tuning of optical and electronic properties, including band edge positions, bandgap energies, spin-orbit interaction strength, and transition intensities.[18,24] While photoluminescence (PL) studies of two-dimensional alloys have been extensive, providing insights into neutral exciton, bound states, and carrier dynamics,[25] relatively little attention has been paid to absorption properties, particularly as a function of temperature and for well-defined monolayers.[26,27] Transmission spectroscopy enables the estimation of absorption spectra and offers a direct probe of the band structure, which can reveal phenomena that may be inaccessible through emission-based techniques, such as higher-energy interband transitions.[28] Additionally, temperature-dependent studies can reveal the strength of electron-phonon interactions and how they evolve across alloys of different compositions.

Performing transmission measurements over a broad temperature range remains technically challenging, especially for atomically thin layers. In this study, we utilize an optical fiber-based configuration, in which mechanically exfoliated monolayers of either $Mo_{1-x}W_xS_2$ or $Mo(S_{1-x}Se_x)_2$ are transferred onto the flat tip of the fiber, covering the entire core region. The fiber with a 2D crystal is placed inside a temperature-controlled cryostat, enabling stable and precise measurements in the range of 20 to 320 K. This allowed us to extract the energies of the interband transitions A, B, and C, and analyze temperature evolution of transition A and B using both Varshni and Bose-Einstein models.[29,30] Moreover, we show that by mixing materials such as $MoS_2$, $WS_2$, and $MoSe_2$, the bandgap can be tuned in the range of approximately 1.6 eV to 2.0 eV at room temperature, and from 1.65 eV to 2.09 eV at cryogenic temperatures. Special attention is also given to the temperature- and composition-dependent splitting between transitions A and B, which provides insights into the variation of spin-orbit interaction strength. Furthermore, we investigate how the choice of transition metal and chalcogen atoms influences the temperature sensitivity of the excitonic transitions and the overall tunability of the optical properties.

**Experimental methods**

Studied materials were obtained from commercially available bulk crystals with known compositions. However, to confirm the composition and crystalline quality of the materials, the $Mo(S_{1-x}Se_x)_2$ and $Mo_{1-x}W_xS_2$ alloys were investigated by X-ray diffraction (XRD) spectroscopy on a Malvern Panalytical Empyrean diffractometer equipped with a copper X-ray tube (λ = 1.540598 Å, $CuK\alpha_1$ radiation). Data were collected in the Bragg–Brentano geometry over a 2θ range of 10° - 80°. For the $Mo(S_{1-x}Se_x)_2$ sample, a fragment of the bulk crystal was placed on a $Si/SiO_2$ substrate and measured over the previously specified scan range. For the $Mo_{1-x}W_xS_2$ alloy, thin flakes were mechanically exfoliated from the bulk crystal, transferred onto the same type of substrate, and then measured. The exfoliation step was necessary because compositional inhomogeneity in the bulk $Mo_{1-x}W_xS_2$ crystal led to ambiguous results when the intact crystal was measured. Lattice parameters were calculated from the (00.6) reflection of $Mo(S_{1-x}Se_x)_2$ and the (00.2) reflection of $Mo_{1-x}W_xS_2$. Assuming a linear dependence of the lattice constant on composition, and taking the lattice parameter of $MoS_2$ as the 0% reference and that of $WS_2$ (or $MoSe_2$) as the 100% reference, the alloy compositions were determined to be 17, 43, 64 and 79% W in $Mo_{1-x}W_xS_2$ and 16, 26, 44, and 55% S in $Mo(S_{1-x}Se_x)_2$, respectively (Figure S1).

The monolayer samples were prepared by mechanical exfoliation and transferred onto the flat end-face of an optical fiber. To improve the adhesion and ensure clean transfer, the fiber tip was gently heated during the process. The fiber with a monolayer was then mounted on the cold finger inside a closed-cycle helium cryostat, allowing for stable temperature control during the experiment. Transmission measurements were carried out by coupling light from a quartz tungsten halogen lamp into the fiber, guiding it through the exfoliated monolayer positioned at the fiber tip, and collecting the transmitted signal with a second fiber aligned at the output. The transmitted spectra were recorded using an optical portable spectrometer. This setup enabled temperature-dependent measurements to be performed across a broad range, from 20 to 320 K.

**Results and discussion**

In this study, we investigated how alloying molybdenum disulfide ($MoS_2$) with either tungsten (cationic alloying) or selenium (anionic alloying) affects its optical properties, as revealed by transmission spectroscopy. Specifically, samples with varying compositions of $Mo_{1-x}W_xS_2$ and $Mo(S_{1-x}Se_x)_2$ were investigated to understand how changes in chemical composition influence the positions of characteristic optical transitions, i.e., A, B, and C excitons. Figures 1b and 1c present the transmission spectra recorded at 20 K for both alloys of selected compositions. For all samples, transitions A and B are clearly resolved, while the high-energy C transition, associated with band nesting,[31] is significantly broader and less pronounced, thus limiting the precision in determining its peak position. This is likely due to its origin away from high-symmetry points of the Brillouin zone, where the band structure is more complex and the density of states is higher.[32–34] The A excitonic transition, corresponding to the direct bandgap at the K point of the Brillouin zone, shifts in energy depending on the material composition, from 1.63 eV in $MoSe_2$ to 2.09 eV in $WS_2$. The simplified schematic of the TMD alloy monolayers discussed here is presented in Figure 1a. A clear trend is observed depending on the type of alloying: chalcogen substitution (Se for S) results in a redshift of the spectral features, whereas metal substitution (W for Mo) leads to a blueshift. This behavior can be understood by considering the atomic properties of the substituting elements. Selenium has a larger atomic radius and lower electronegativity than sulfur, leading to weaker orbital overlap and a reduced crystal field strength around the transition metal, which, in turn, lowers the bandgap and results in a redshift of optical transitions.[35,36] In contrast, although molybdenum and tungsten have similar atomic radii, tungsten has a higher atomic number and significantly stronger spin–orbit coupling. This leads to greater splitting of the valence bands and a modified electronic band structure that increases the bandgap energy, manifesting as a blueshift in the optical spectra.[37] These trends are illustrated in Figure 1d, which shows the energies of A and B excitonic transitions as a function of alloy composition. The extracted bowing parameters for the A excitonic transition were $b_A=0.024\pm0.014$[38] for the $Mo(S_{1-x}Se_x)_2$ alloy and $b_A=0.17\pm0.05$[39,40] for the $Mo_{1-x}W_xS_2$ system, while for the B exciton, the bowing values were $b_B=0.018\pm0.012$[38] and $b_B=0.054\pm0.037$[25], respectively. These values are consistent with previously reported data for similar alloy systems, indicating that the variation in excitonic transition energies with composition follows nonlinear behavior. However, it is important to note that in both cases, the bowing parameters are close to zero, which implies structural and electronic similarity between the alloy constituents. Moreover, the energy separation between transitions A and B, corresponding to the spin-orbit splitting of the valence band, is also considered, as shown in Figure 1e. The smallest A–B splitting is observed for pure $MoS_2$ and increases with both chalcogen and metal alloying. However, the effect is significantly stronger for metal substitution: for $WS_2$, the splitting reaches up to 0.4 eV. This observation is consistent with theoretical expectations, as tungsten atoms exhibit stronger spin–orbit coupling than molybdenum, resulting in greater splitting of the valence bands at the K point.[41] In contrast, the substitution of sulfur with selenium also influences the spin–orbit interaction, but to a lesser extent due to the smaller difference in atomic mass and spin–orbit coupling strength between S and Se.[23]

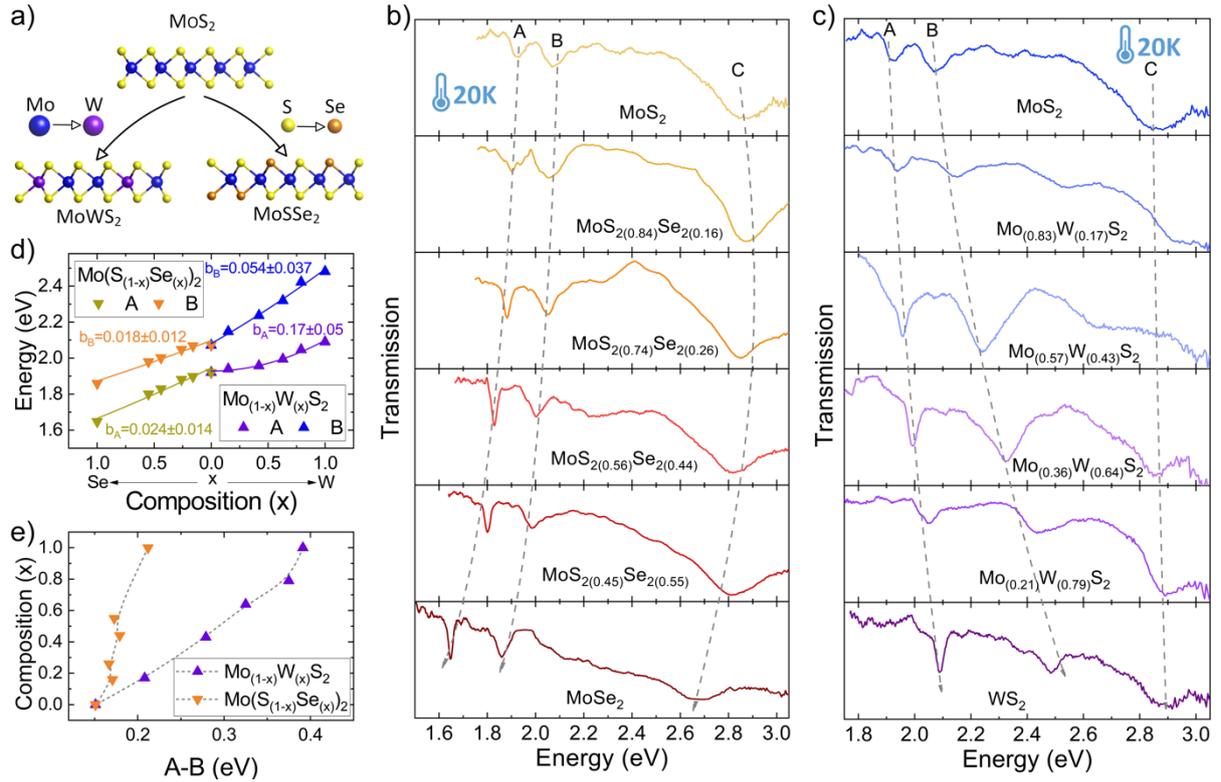

Figure 1. (a) Schematic representation of the studied TMD alloy monolayers: $Mo_{1-x}W_xS_2$, with blue Mo, yellow S, and purple W atoms; and $Mo(S_{1-x}Se_x)_2$, with orange Se atoms. Transmission spectra of (b) $Mo_{1-x}W_xS_2$ and (c) $Mo(S_{1-x}Se_x)_2$ samples measured at 20 K. (d) Composition-dependent energies of A and B excitons for $Mo_{1-x}W_xS_2$ and $Mo(S_{1-x}Se_x)_2$. (e) Composition-dependent energy splitting between A and B transitions for $Mo_{1-x}W_xS_2$ and $Mo(S_{1-x}Se_x)_2$.

As described earlier, a monolayer of each alloy composition was transferred onto the flat-cleaved facet of a multimode optical fiber and placed in a closed-cycle cryostat. To study the impact of external stimuli, transmission spectra were recorded over a wide temperature range from 320 down to 20 K. Representative results are presented in Figure 2 for the $Mo(S_{1-x}Se_x)_2$ alloy series, from pure $MoS_2$ through intermediate compositions to pure $MoSe_2$. A systematic blueshift of transitions A, B, and C is observed as the temperature decreases. For most compositions, transitions A and B remain clearly resolved up to room temperature, allowing reliable peak extraction across the entire range. In contrast, the C transition appears significantly broadened and exhibits minimal temperature-dependent shift. Due to the increased uncertainty in determining its peak position, this feature was excluded from further quantitative analysis.

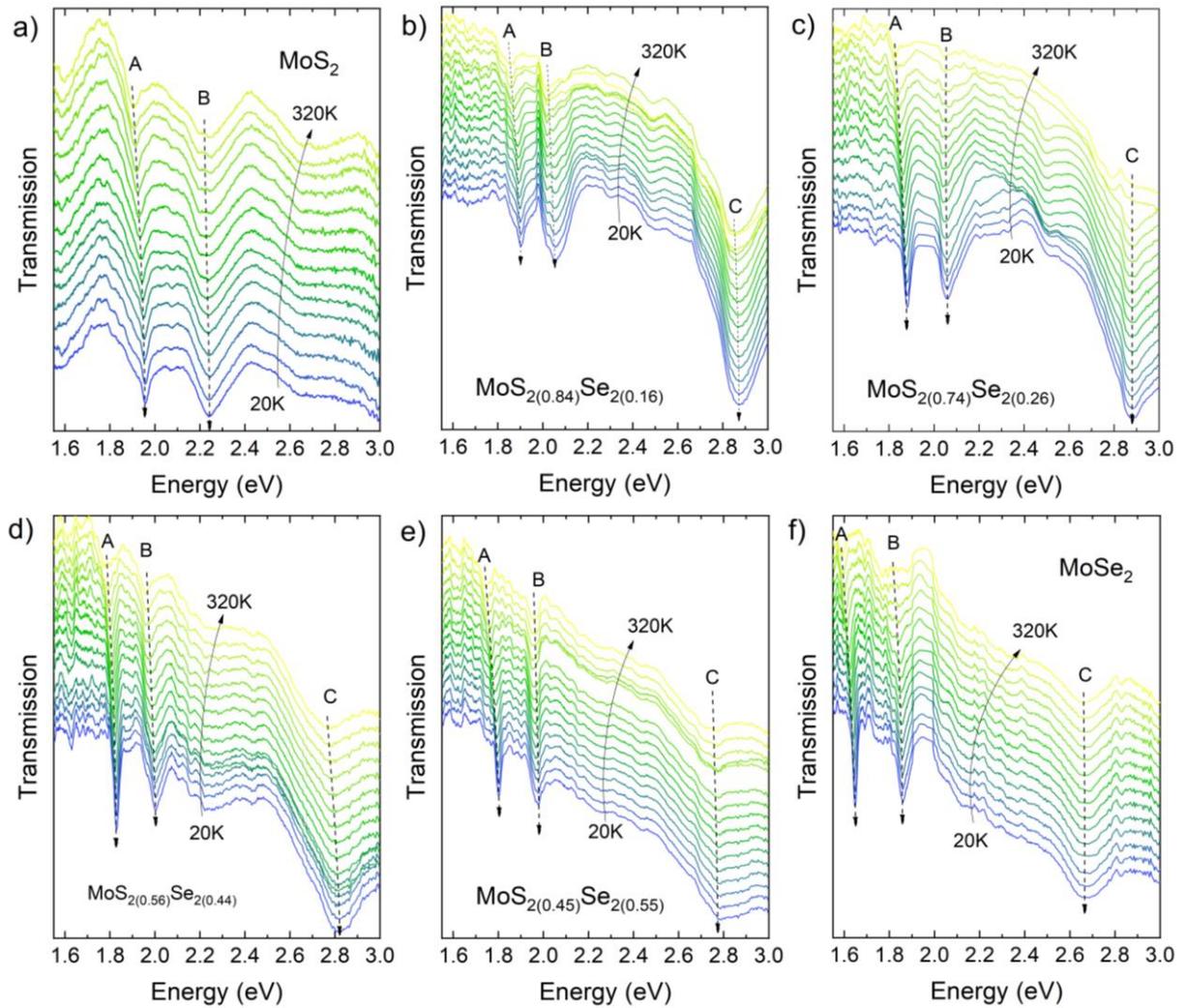

Figure 2. Transmission spectra of Mo(S$_{1-x}$Se$_x$)$_2$ alloys with varying selenium content, measured in the temperature range 20–320 K. Panels (a)–(f) correspond to increasing Se concentration. A clear shift of the transmission minimum with increasing temperature and Se content is observed, reflecting changes in the optical transition energies of the studied materials.

Furthermore, Figure 3 shows the analogous results for Mo$_{1-x}$W$_x$S$_2$ alloy monolayers, following the same trend. A pronounced blueshift of the A and B excitonic transitions is again observed as temperature decreases. The energy of excitonic transitions was extracted using a Gaussian fitting procedure applied to each peak. As expected from excitonic behavior, the amplitude of the features increases at lower temperatures, while their linewidths decrease due to reduced electron–phonon interaction strength and a lower phonon population.

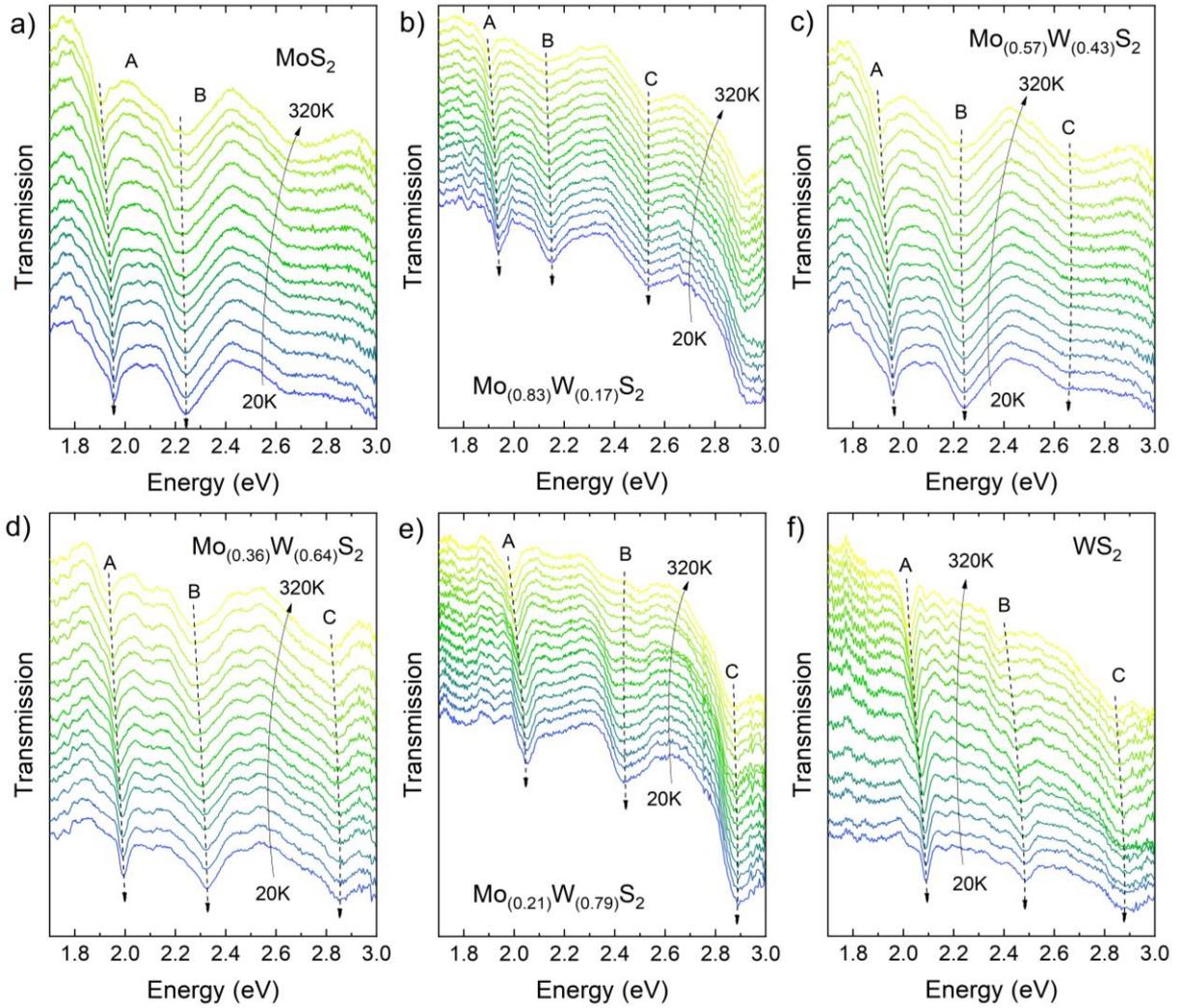

Figure 3. Transmission spectra of $Mo_{1-x}W_xS_2$ alloys with varying selenium content, measured in the temperature range 20–320 K. Panels (a)–(f) correspond to increasing W concentration. A clear shift of the transmission minimum with increasing temperature and W content is observed, reflecting changes in the optical transition energies of the studied materials.

The extracted A and B exciton energies as a function of temperature are plotted in Figure 4 for all compositions and for both alloys. For the A transition in the $Mo(S_{1-x}Se_x)_2$ series, we observe that the rate of energy decrease with increasing temperature is directly correlated with the chalcogen content. It can be seen that pure $MoS_2$ shows the smallest thermal shift, and $MoSe_2$ the largest, which is consistent with previous studies.[8] Similarly, the $Mo_{1-x}W_xS_2$ alloy exhibits a systematic evolution: $WS_2$ undergoes the strongest thermal shift, and this effect gradually diminishes toward $MoS_2$ as the W content decreases. In all investigated materials trend is more pronounced for transition A than for B, where the incorporation of tungsten increases the temperature sensitivity of the B exciton to a lesser extent. This suggests that the A exciton is more sensitive to changes in the electronic band structure and exciton-phonon coupling introduced by metal substitution.

To quantitatively describe the temperature dependence of the excitonic transitions, the data were fitted using both the Varshni model (Figure 4) and a Bose-Einstein-type model. The Varshni equation is an empirical model commonly used to describe the temperature-induced bandgap shrinkage:

$$E_g(T) = E_0 - \frac{\alpha T^2}{T+\beta} \qquad (1)$$

where $E_g(T)$ is the transition energy at temperature T, $E_0$ is the energy at 0 K, and α is a material-specific coefficient that quantifies the strength of the bandgap reduction with temperature related to electron-phonon coupling and thermal expansion effects; $\beta$ is a parameter that can be interpreted as an effective temperature scale at which the bandgap energy changes more rapidly.[29] Additionally, we employed a Bose-Einstein model that accounts more explicitly for exciton-phonon interactions:

$$E_g(T) = E_0 - \frac{a_B}{\exp\left(\frac{\theta_B}{T}\right)-1} \qquad (2)$$

Where $a_B$ is a coupling constant that represents the strength of the exciton–phonon interaction, and $\theta_B$ is an effective phonon temperature and is related to the average energy of the phonons involved in the interaction.[30] Both models were applied to the extracted A and B transition energies across the whole temperature range. The resulting fitting parameters are summarized in Table 1.

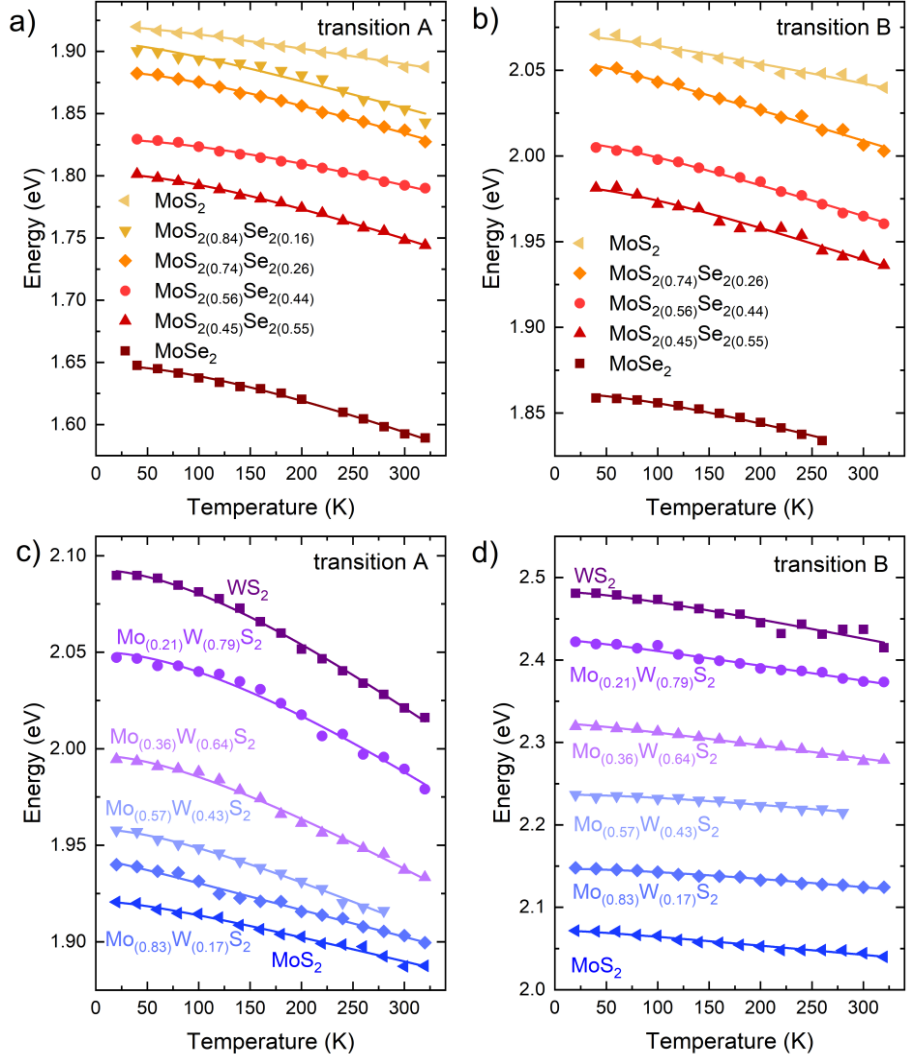

Figure 4. The temperature dependence of the excitonic transition energies, along with the corresponding Varshni fits, reveals how the thermal sensitivity of these materials varies systematically with composition: in the Mo(S$_{1-x}$Se$_x$)$_2$ alloy, transitions A (a) and B (b), and in Mo$_{1-x}$W$_x$S$_2$ transitions A (c) and B (d).

These observations are also illustrated in Figure 5, which shows the total energy shifts of the A and B excitonic transitions as a function of temperature for all the studied materials. For the A transition in pure MoS$_2$, the shift is only about 35 meV, indicating low sensitivity to temperature changes, consistent with expectations and the trends discussed earlier for this material. In contrast, the B transitions exhibit more variability, possibly due to how the lower valence band (split by spin-orbit interaction) interacts with phonons or responds to alloy disorder.[42] For the Mo(S$_{1-x}$Se$_x$)$_2$ alloys, the A transition shifts range from 35 to 60 meV, while the B transitions span from 10 to 47 meV. This reflects a higher temperature sensitivity compared to pure $MoS_2$, particularly for the A exciton. The $Mo_{1-x}W_xS_2$ alloys show even more pronounced temperature-induced shifts: from 35 to 75 meV for the A transition, and from 10 to 67 meV for the B transition when Mo is replaced with W. These large shifts indicate stronger exciton–phonon interactions, likely due to both the heavier atomic mass of W compared to Mo (affecting optical phonon modes) and more delocalized electronic states that enhance coupling with lattice vibrations.[43,44] The increasing range of energy shifts with higher W or Se content suggests that the temperature sensitivity of the bandgap can be tuned through composition, which is relevant for designing optoelectronic devices operating under variable thermal conditions.

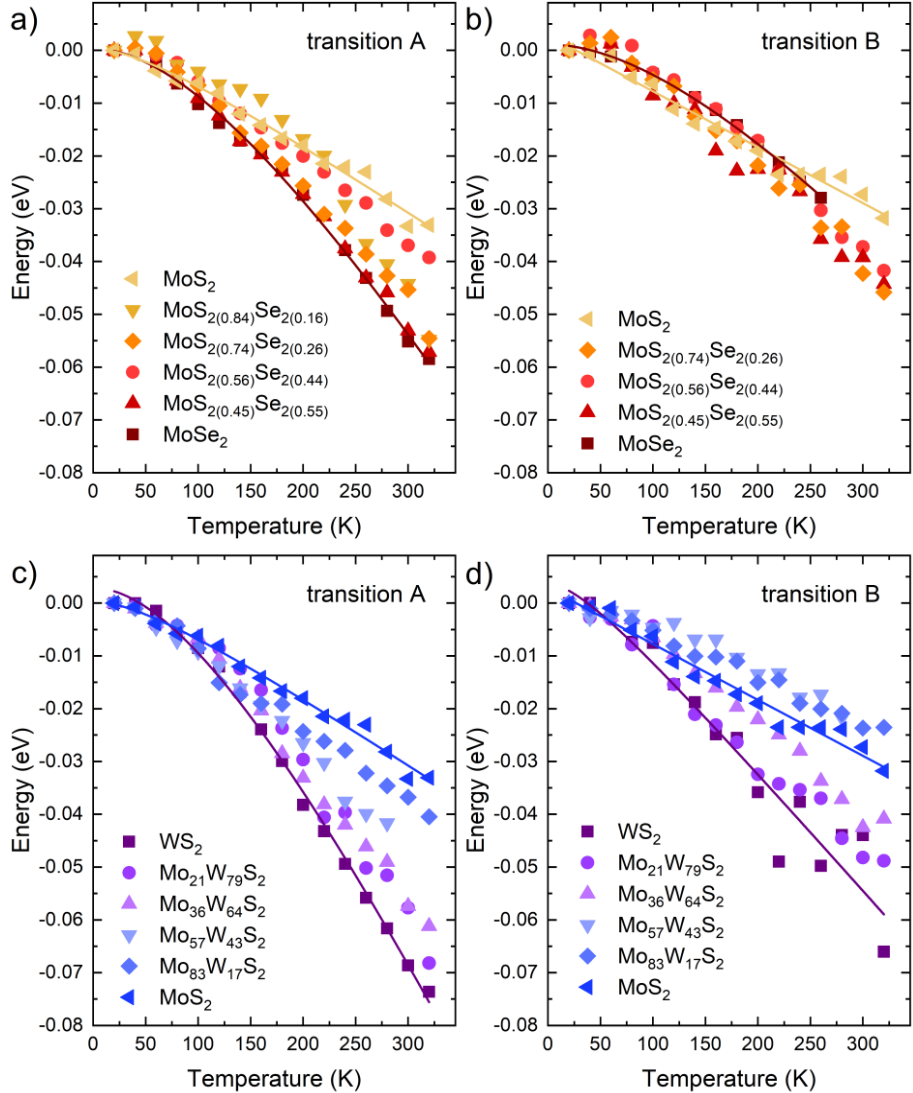

Figure 5. Total energy shift of the A and B excitonic transitions between 20 and 320 K for $Mo(S_{1-x}Se_x)_2$ and $Mo_{1-x}W_xS_2$ alloys.

In the case of the A transition in $Mo_{1-x}W_xS_2$ alloys, both $\alpha$ and $a_B$ parameters increase significantly with increasing W content. This indicates that the sensitivity of exciton energy to temperature is stronger in $WS_2$-rich samples and gradually weakens as the composition shifts toward $MoS_2$. These trends show that $MoS_2$ has a more thermally stable bandgap, which, considering its less stiff lattice compared to $WS_2$,[45] further supports the dominant role of exciton-phonon coupling in thermally induced bandgap changes. Moreover, the $Mo(S_{1-x}Se_x)_2$ alloys, where the chalcogen atom is gradually substituted from S to Se, also display systematic behavior. The Se-rich alloy shows the highest values of $\alpha$ and $a_B$, indicating strong exciton-phonon interaction and significant energy shifts with temperature, which decrease monotonically for the intermediate compositions of the ternary alloy. However, when considering both alloy systems, the temperature-induced energy shift is greater upon W substitution, indicating that metal substitution has a stronger impact on the lattice structure and exciton-phonon coupling than chalcogen substitution, at least in the context of temperature-dependent optical properties.

Table 1. Varshni and Bose-Einstein fitting results for the temperature-induced energy shifts of A and B excitons.

| Material | transition | $E_0$ (eV) | $\alpha$ ($10^{-4}$ eV/K) | $E_0$ (eV) | $a_B$ (meV) |
|---|---|---|---|---|---|
| $MoS_2$ | A | 1.9207(15) | 1.33(30) | 1.92222(64) | 2.89(2.0) |
|  | B | 2.0706(11) | 1.26(15) | 2.071(94) | 5.6(1.6) |
| $MoS_{2(0.84)}Se_{2(16)}$ | A | 1.9073(15) | 2.34(30) | 1.9034(17) | 15.1(2.0) |
| $MoS_{2(0.74)}Se_{2(0.26)}$ | A | 1.8845(54) | 2.68(37) | 1.8829(13) | 20.6(4.0) |
|  | B | 2.0560(10) | 1.826(62) | 2.05335(92) | 8.93(94) |
| $MoS_{2(0.56)}Se_{2(0.44)}$ | A | 1.8297(49) | 2.49(22) | 1.83031(74) | 17.2(2.7) |
|  | B | 2.0089(64) | 1.949(46) | 2.0082(74) | 9.56(85) |
| $MoS_{2(0.45)}Se_{2(0.55)}$ | A | 1.8022(77) | 3.12(29) | 1.80016(98) | 26.7(4.3) |
|  | B | 1.9826(12) | 2.152(42) | 1.9830(12) | 10.6(1.4) |
| $MoSe_2$ | A | 1.6480(11) | 3.62(56) | 1.6460(13) | 30.6(6.2) |
|  | B | 1.8606(39) | 1.85(55) | 1.86090(30) | 12(3) |
| $Mo_{(0.83)}W_{(0.17)}S_2$ | A | 1.9425(25) | 1.40(34) | 1.9413(67) | 3.51(4.85) |
|  | B | 2.1469(48) | 1.247(96) | 2.14778(41) | 6.06(76) |
| $Mo_{(0.57)}W_{(0.43)}S_2$ | A | 1.95827(87) | 2.34(27) | 1.9572(95) | 15.0(3.5) |
|  | B | 2.2343(67) | 1.38(22) | 2.2354(11) | 7.17(60) |
| $Mo_{(0.36)}W_{(0.64)}S_2$ | A | 1.9965(13) | 3.17(41) | 1.9942(11) | 28.1(4.7) |
|  | B | 2.3231(73) | 1.655(92) | 2.3217(67) | 8.14(88) |
| $Mo_{(0.21)}W_{(0.79)}S_2$ | A | 2.0502(12) | 4.15(26) | 2.0420(11) | 36.7(1.2) |
|  | B | 2.4247(31) | 1.85(38) | 2.4222(12) | 9.4(1.5) |
| $WS_2$ | A | 2.09269(96) | 4.43(48) | 2.09003(59) | 40.6(3.2) |
|  | B | 2.4828(25) | 2.52(38) | 2.4829(25) | 11.1(1.6) |

To better visualize the temperature-dependent behavior of the excitonic transitions, both the $\alpha$ and $a_B$ parameters are presented graphically in Figure 6, where the left Y-axis corresponds to the values obtained from the Varshni model, while the right Y-axis reflects those from the Bose-Einstein model. As shown, both models reveal the same qualitative trends for each material system, confirming their consistency in describing the exciton-phonon interaction strength. Although the temperature dependence of excitonic transitions in TMDCs has been extensively studied in bulk crystals, the behavior observed in monolayers is distinct due to their reduced dimensionality. In monolayer systems, both the Varshni and Bose–Einstein parameters tend to be systematically lower than in bulk counterparts, reflecting weaker exciton–phonon coupling. Our results confirm this trend across all studied compounds. The temperature-induced redshift of excitonic transitions is less pronounced in monolayers, and the associated parameters show a clear dependence on material composition. In $Mo_{1-x}W_xS_2$ alloys, increasing tungsten content leads to higher sensitivity to temperature, with the most significant changes occurring in $WS_2$-rich samples. This suggests that lattice composition can partially compensate for the dimensional reduction by enhancing coupling to specific phonon modes.

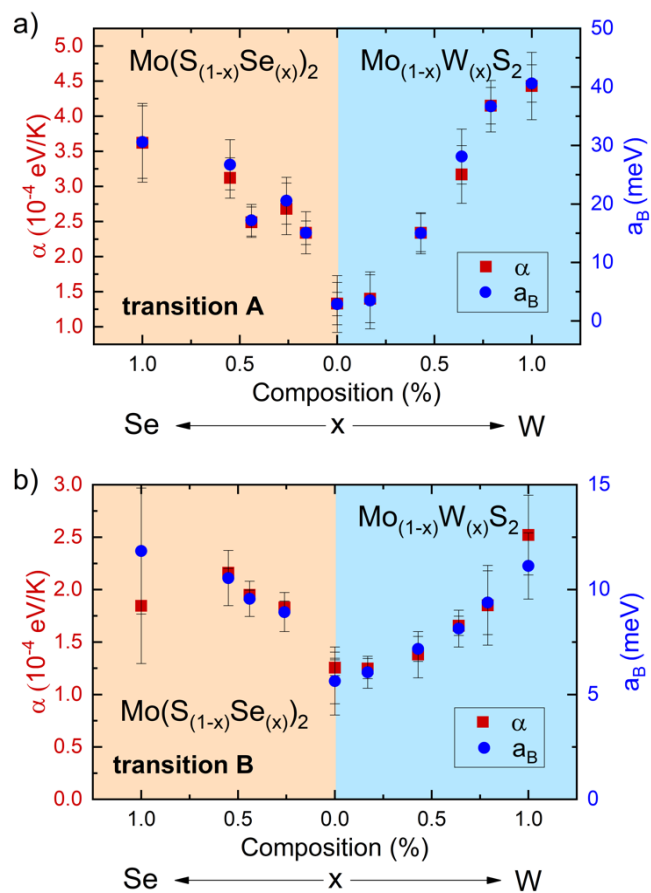

Figure 6. Comparison of the parameters $\alpha$ (Varshni model) and $a_B$ (Bose–Einstein model) for different compositions of $Mo_{1-x}W_xS_2$ and $Mo(S_{1-x}Se_x)_2$.

## Summary

In this article, we investigated the shift of the fundamental excitonic transitions A, B, and C in monolayer alloys derived from $MoS_2$ as the base material. Introducing selenium (Se) on a chalcogen site leads to a redshift of the transitions (i.e., lower energy), while substituting molybdenum (Mo) with tungsten (W) on a transition metal site causes a blueshift (i.e., higher energy). These trends provide valuable insights into how the chemical composition can be tuned to achieve excitonic transitions at desired wavelengths in a controlled and stable manner. Another key aspect of our study was the investigation of temperature-dependent behavior. It was found that replacing sulfur with selenium increases the material's sensitivity to temperature. On the other hand, substituting Mo with W results in a stronger temperature dependence, likely due to the heavier atomic mass of tungsten. In this case, the energy shifts exhibit more monotonic behavior with respect to the alloy composition. These findings highlight the importance of composition engineering for both spectral and thermal control in transition metal dichalcogenide alloys.

## Conflict of Interest Statement

The authors have no conflicts to disclose.

## Acknowledgments

This work was supported by the National Science Centre (NCN) Poland OPUS 23 no. 2022/45/B/ST7/02750.

## Data availability

The data that support the findings of this study are openly available in Zenodo, at https://doi.org/10.5281/zenodo.16632871, reference number 10.5281/zenodo.16632871.

References

(1) A Review of the Synthesis, Properties, and Applications of 2D Materials - Shanmugam - 2022 - Particle & Particle Systems Characterization - Wiley Online Library. https://onlinelibrary.wiley.com/doi/full/10.1002/ppsc.202200031 (accessed 2024-12-30).

(2) Dutta, T.; Yadav, N.; Wu, Y.; Cheng, G. J.; Liang, X.; Ramakrishna, S.; Sbai, A.; Gupta, R.; Mondal, A.; Hongyu, Z.; Yadav, A. Electronic Properties of 2D Materials and Their Junctions. Nano Materials Science **2023**. https://doi.org/10.1016/j.nanoms.2023.05.003.

(3) Hu, J.-Q.; Shi, X.-H.; Wu, S.-Q.; Ho, K.-M.; Zhu, Z.-Z. Dependence of Electronic and Optical Properties of $MoS_2$ Multilayers on the Interlayer Coupling and Van Hove Singularity. Nanoscale Res Lett **2019**, 14 (1), 288. https://doi.org/10.1186/s11671-019-3105-9.

(4) Zheng, W.; Jiang, Y.; Hu, X.; Li, H.; Zeng, Z.; Wang, X.; Pan, A. Light Emission Properties of 2D Transition Metal Dichalcogenides: Fundamentals and Applications. Advanced Optical Materials **2018**, 6 (21), 1800420. https://doi.org/10.1002/adom.201800420.

(5) Jin, L.; Wang, H.; Cao, R.; Khan, K.; Tareen, A. K.; Wageh, S.; Al-Ghamdi, A.; Li, S.; Li, D.; Zhang, Y. The Rise of 2D Materials/Ferroelectrics for next Generation Photonics and Optoelectronics Devices. APL Materials **2022**, 10, 60903. https://doi.org/10.1063/5.0094965.

(6) All-Optical Modulation Technology Based on 2D Layered Materials. https://www.mdpi.com/2072-666X/13/1/92 (accessed 2025-04-16).

(7) Kumbhakar, P.; Chowde Gowda, C.; Tiwary, C. S. Advance Optical Properties and Emerging Applications of 2D Materials. Frontiers in Materials **2021**, 8.

(8) Kopaczek, J.; Zelewski, S.; Yumigeta, K.; Sailus, R.; Tongay, S.; Kudrawiec, R. Temperature Dependence of the Indirect Gap and the Direct Optical Transitions at the High-Symmetry Point of the Brillouin Zone and Band Nesting in $MoS_2$, $MoSe_2$, $MoTe_2$, $WS_2$, and $WSe_2$ Crystals. J. Phys. Chem. C **2022**, 126 (12), 5665–5674. https://doi.org/10.1021/acs.jpcc.2c01044.

(9) Santra, P.; Ghaderzadeh, S.; Ghorbani-Asl, M.; Komsa, H.-P.; Besley, E.; Krasheninnikov, A. V. Strain-Modulated Defect Engineering of Two-Dimensional Materials. npj 2D Mater Appl **2024**, 8 (1), 1–9. https://doi.org/10.1038/s41699-024-00472-x.

(10) Liang, Q.; Zhang, Q.; Zhao, X.; Liu, M.; Wee, A. T. S. Defect Engineering of Two-Dimensional Transition-Metal Dichalcogenides: Applications, Challenges, and Opportunities. ACS Nano **2021**, 15 (2), 2165–2181. https://doi.org/10.1021/acsnano.0c09666.

(11) Hossen, M. F.; Shendokar, S.; Aravamudhan, S. Defects and Defect Engineering of Two-Dimensional Transition Metal Dichalcogenide (2D TMDC) Materials. Nanomaterials **2024**, 14 (5), 410. https://doi.org/10.3390/nano14050410.

(12) Lin, Z.; Carvalho, B.; Kahn, E.; Lv, R.; Rao, R.; Terrones, H.; Pimenta, M.; Terrones, M. Defect Engineering of Two-Dimensional Transition Metal Dichalcogenides. 2D Materials **2016**, 3, 022002. https://doi.org/10.1088/2053-1583/3/2/022002.

(13) Lin, Y.-C.; Torsi, R.; Geohegan, D. B.; Robinson, J. A.; Xiao, K. Controllable Thin-Film Approaches for Doping and Alloying Transition Metal Dichalcogenides Monolayers. Advanced Science **2021**, 8 (9), 2004249. https://doi.org/10.1002/advs.202004249.

(14) Baithi, M.; Duong, D. L. Doped, Two-Dimensional, Semiconducting Transition Metal Dichalcogenides in Low-Concentration Regime. Crystals **2024**, 14 (10), 832. https://doi.org/10.3390/cryst14100832.

(15) Lan, C.; Li, C.; Ho, J.; Liu, Y. 2D $WS_2$: From Vapor Phase Synthesis to Device Applications. Advanced Electronic Materials **2020**, 7. https://doi.org/10.1002/aelm.202000688.

(16)     Niu, Y.; Gonzalez-Abad, S.; Frisenda, R.; Marauhn, P.; Drüppel, M.; Gant, P.; Schmidt, R.; Taghavi, N. S.; Barcons, D.; Molina-Mendoza, A. J.; De Vasconcellos, S. M.; Bratschitsch, R.; Perez De Lara, D.; Rohlfing, M.; Castellanos-Gomez, A. Thickness-Dependent Differential Reflectance Spectra of Monolayer and Few-Layer MoS2, MoSe2, WS2 and WSe2. Nanomaterials **2018**, 8 (9), 725. https://doi.org/10.3390/nano8090725.

(17)     Wang, G.; Robert, C.; Suslu, A.; Chen, B.; Yang, S.; Alamdari, S.; Gerber, I. C.; Amand, T.; Marie, X.; Tongay, S.; Urbaszek, B. Spin-Orbit Engineering in Transition Metal Dichalcogenide Alloy Monolayers. Nat Commun **2015**, 6 (1), 10110. https://doi.org/10.1038/ncomms10110.

(18)     Sulfur Vacancy Related Optical Transitions in Graded Alloys of MoxW1-xS2 Monolayers - Ghafariasl - 2024 - Advanced Optical Materials - Wiley Online Library. https://advanced.onlinelibrary.wiley.com/doi/full/10.1002/adom.202302326 (accessed 2025-04-16).

(19)     Krishnan, U.; Kaur, M.; Singh, K.; Kumar, M.; Kumar, A. A Synoptic Review of MoS2: Synthesis to Applications. Superlattices and Microstructures **2019**, 128, 274–297. https://doi.org/10.1016/j.spmi.2019.02.005.

(20)     Splendiani, A.; Sun, L.; Zhang, Y.; Li, T.; Kim, J.; Chim, C.-Y.; Galli, G.; Wang, F. Emerging Photoluminescence in Monolayer MoS2. Nano Lett. **2010**, 10 (4), 1271–1275. https://doi.org/10.1021/nl903868w.

(21)     Kopaczek, J.; Polak, M. P.; Scharoch, P.; Wu, K.; Chen, B.; Tongay, S.; Kudrawiec, R. Direct Optical Transitions at K- and H-Point of Brillouin Zone in Bulk MoS2, MoSe2, WS2, and WSe2. Journal of Applied Physics **2016**, 119 (23), 235705. https://doi.org/10.1063/1.4954157.

(22)     Mak, K. F.; Lee, C.; Hone, J.; Shan, J.; Heinz, T. F. Atomically Thin ${\mathrm{MoS}}_{2}$: A New Direct-Gap Semiconductor. Phys. Rev. Lett. **2010**, 105 (13), 136805. https://doi.org/10.1103/PhysRevLett.105.136805.

(23)     Zhu, Z. Y.; Cheng, Y. C.; Schwingenschlögl, U. Giant Spin-Orbit-Induced Spin Splitting in Two-Dimensional Transition-Metal Dichalcogenide Semiconductors. Phys. Rev. B **2011**, 84 (15), 153402. https://doi.org/10.1103/PhysRevB.84.153402.

(24)     Zhao, W.; Zheng, T.; Cui, Y.; Song, J.; Liu, H.; Lu, J.; Ni, Z. Observation of Band Gap Bowing Effect Vanishing in Graded-Composition Monolayer Mo1−xWxS2 Alloy. Applied Physics Letters **2024**, 124 (7), 073102. https://doi.org/10.1063/5.0188793.

(25)     Tunable Band Gap Photoluminescence from Atomically Thin Transition-Metal Dichalcogenide Alloys | ACS Nano. https://pubs.acs.org/doi/abs/10.1021/nn401420h (accessed 2025-04-16).

(26)     Chang, Y.-C.; Wang, Y.-K.; Chen, Y.-T.; Der Yuh, L. Facile and Reliable Thickness Identification of Atomically Thin Dichalcogenide Semiconductors Using Hyperspectral Microscopy. Nanomaterials **2020**, 10, 526. https://doi.org/10.3390/nano10030526.

(27)     Frisenda, R.; Niu, Y.; Gant, P.; Molina-Mendoza, A.; Schmidt, R.; Bratschitsch, R.; Liu, J.; Fu, L.; Dumcenco, D.; Kis, A.; Perez de Lara, D.; Castellanos-Gomez, A. Micro-Reflectance and Transmittance Spectroscopy: A Versatile and Powerful Tool to Characterize 2D Materials. Journal of Physics D: Applied Physics **2017**, 50. https://doi.org/10.1088/1361-6463/aa5256.

(28)     Kudrawiec, R.; Walukiewicz, W. Electromodulation Spectroscopy of Highly Mismatched Alloys. Journal of Applied Physics **2019**, 126 (14), 141102. https://doi.org/10.1063/1.5111965.

(29)     Varshni, Y. P. Temperature Dependence of the Energy Gap in Semiconductors. Physica **1967**, 34 (1), 149–154. https://doi.org/10.1016/0031-8914(67)90062-6.

(30)     Viña, L.; Logothetidis, S.; Cardona, M. Temperature Dependence of the Dielectric Function of Germanium. Phys. Rev. B **1984**, 30 (4), 1979–1991. https://doi.org/10.1103/PhysRevB.30.1979.

(31) Kozawa, D.; Kumar, R.; Carvalho, A.; Kumar Amara, K.; Zhao, W.; Wang, S.; Toh, M.; Ribeiro, R. M.; Castro Neto, A. H.; Matsuda, K.; Eda, G. Photocarrier Relaxation Pathway in Two-Dimensional Semiconducting Transition Metal Dichalcogenides. Nat Commun **2014**, 5 (1), 4543. https://doi.org/10.1038/ncomms5543.

(32) Kopaczek, J.; Woźniak, T.; Tamulewicz-Szwajkowska, M.; Zelewski, S. J.; Serafińczuk, J.; Scharoch, P.; Kudrawiec, R. Experimental and Theoretical Studies of the Electronic Band Structure of Bulk and Atomically Thin Mo$_{1-x}$W$_x$Se$_2$ Alloys. ACS Omega **2021**, 6 (30), 19893–19900. https://doi.org/10.1021/acsomega.1c02788.

(33) Biroju, R. K.; Maity, D.; Vretenár, V.; Vančo, Ľ.; Sharma, R.; Sahoo, M. R.; Kumar, J.; Gayathri, G.; Narayanan, T. N.; Nayak, S. K. Quantification of Alloy Atomic Composition Sites in 2D Ternary MoS2(1-x)Se2x and Their Role in Persistent Photoconductivity, Enhanced Photoresponse and Photo-Electrocatalysis. Materials Today Advances **2024**, 22, 100504. https://doi.org/10.1016/j.mtadv.2024.100504.

(34) Hong, J.; Senga, R.; Pichler, T.; Suenaga, K. Probing Exciton Dispersions of Freestanding Monolayer $\mathrm{WSe}_2$ by Momentum-Resolved Electron Energy-Loss Spectroscopy. Phys. Rev. Lett. **2020**, 124 (8), 087401. https://doi.org/10.1103/PhysRevLett.124.087401.

(35) Gong, Y.; Liu, Z.; Lupini, A. R.; Shi, G.; Lin, J.; Najmaei, S.; Lin, Z.; Elías, A. L.; Berkdemir, A.; You, G.; Terrones, H.; Terrones, M.; Vajtai, R.; Pantelides, S. T.; Pennycook, S. J.; Lou, J.; Zhou, W.; Ajayan, P. M. Band Gap Engineering and Layer-by-Layer Mapping of Selenium-Doped Molybdenum Disulfide. Nano Lett. **2014**, 14 (2), 442–449. https://doi.org/10.1021/nl4032296.

(36) Zhao, W.; Ghorannevis, Z.; Chu, L.; Toh, M.; Kloc, C.; Tan, P.-H.; Eda, G. Evolution of Electronic Structure in Atomically Thin Sheets of WS2 and WSe2. ACS Nano **2013**, 7 (1), 791–797. https://doi.org/10.1021/nn305275h.

(37) Kormányos, A.; Burkard, G.; Gmitra, M.; Fabian, J.; Zólyomi, V.; Drummond, N. D.; Fal'ko, V. K·p Theory for Two-Dimensional Transition Metal Dichalcogenide Semiconductors. 2D Mater. **2015**, 2 (2), 022001. https://doi.org/10.1088/2053-1583/2/2/022001.

(38) Feng, Q.; Mao, N.; Wu, J.; Xu, H.; Wang, C.; Zhang, J.; Xie, L. Growth of MoS2(1–x)Se2x (x = 0.41–1.00) Monolayer Alloys with Controlled Morphology by Physical Vapor Deposition. ACS Nano **2015**, 9 (7), 7450–7455. https://doi.org/10.1021/acsnano.5b02506.

(39) Kim, J.; Seung, H.; Kang, D.; Kim, J.; Bae, H.; Park, H.; Kang, S.; Choi, C.; Choi, B. K.; Kim, J. S.; Hyeon, T.; Lee, H.; Kim, D.-H.; Shim, S.; Park, J. Wafer-Scale Production of Transition Metal Dichalcogenides and Alloy Monolayers by Nanocrystal Conversion for Large-Scale Ultrathin Flexible Electronics. Nano Lett. **2021**, 21 (21), 9153–9163. https://doi.org/10.1021/acs.nanolett.1c02991.

(40) Song, J.-G.; Ryu, G. H.; Lee, S. J.; Sim, S.; Lee, C. W.; Choi, T.; Jung, H.; Kim, Y.; Lee, Z.; Myoung, J.-M.; Dussarrat, C.; Lansalot-Matras, C.; Park, J.; Choi, H.; Kim, H. Controllable Synthesis of Molybdenum Tungsten Disulfide Alloy for Vertically Composition-Controlled Multilayer. Nat Commun **2015**, 6, 7817. https://doi.org/10.1038/ncomms8817.

(41) Lee, Y.; Eu, P.; Lim, C.; Cha, J.; Kim, S.; Denlinger, J. D.; Kim, Y. Controlling Spin-Orbit Coupling Strength of Bulk Transition Metal Dichalcogenide Semiconductors. Current Applied Physics **2021**, 30, 4–7. https://doi.org/10.1016/j.cap.2021.03.008.

(42) Zhang, Y.; Li, H.; Wang, H.; Liu, R.; Zhang, S.-L.; Qiu, Z.-J. On Valence-Band Splitting in Layered MoS2. ACS Nano **2015**, 9 (8), 8514–8519. https://doi.org/10.1021/acsnano.5b03505.

(43) Qiu, D. Y.; da Jornada, F. H.; Louie, S. G. Optical Spectrum of $\mathrm{MoS}_2$: Many-Body Effects and Diversity of Exciton States. Phys. Rev. Lett. **2013**, 111 (21), 216805. https://doi.org/10.1103/PhysRevLett.111.216805.

(44) Molina-Sánchez, A.; Wirtz, L. Phonons in Single-Layer and Few-Layer MoS$_2$ and WS$_2$. Phys. Rev. B **2011**, 84 (15), 155413. https://doi.org/10.1103/PhysRevB.84.155413.

(45) Bandgap engineering of MoS2/MX2 (MX2 = WS2, MoSe2 and WSe2) heterobilayers subjected to biaxial strain and normal compressive strain - RSC Advances (RSC Publishing). https://pubs.rsc.org/en/content/articlelanding/2016/ra/c5ra27871f?utm_source=chatgpt.com (accessed 2025-06-09).

(46) Ho, C. H.; Liao, P. C.; Huang, Y. S.; Tiong, K. K. Temperature Dependence of Energies and Broadening Parameters of the Band-Edge Excitons of ReS2 and ReSe2. Phys. Rev. B **1997**, 55 (23), 15608–15613. https://doi.org/10.1103/PhysRevB.55.15608.

**Supporting information file for:**

# Sensitivity of excitonic transitions to temperature in monolayers of TMD alloys

K. Ciesiołkiewicz-Klepek[1, a)], J. Kopaczek[1], J. Serafińczuk[2], R. Kudrawiec[1, b)]

*[1]Department of Semiconductor Materials Engineering,*
*Wroclaw University of Science and Technology,*
*Wybrzeże Wyspiańskiego 27, 50-370 Wrocław, Poland*

*[2]Department of Nanometrology,*
*Wroclaw University of Science and Technology,*
*Wybrzeże Wyspiańskiego 27, 50-370 Wrocław, Poland*

[a)] e-mail address: karolina.ciesiolkiewicz@pwr.edu.pl
[b)] e-mail address: robert.kudrawiec@pwr.edu.pl

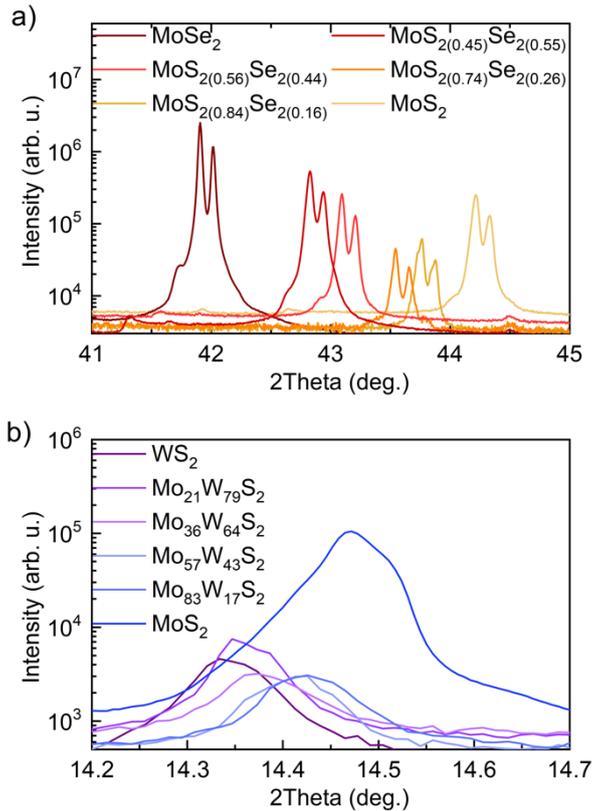

Figure S1. X-ray diffraction (XRD) patterns of $Mo(S_{1-x}Se_x)_2$ (a) and $Mo_{1-x}W_xS_2$ (b) alloys measured using $CuK\alpha_1$ radiation ($\lambda$ = 1.540598 Å) in Bragg–Brentano geometry. For $Mo(S_{1-x}Se_x)_2$, bulk fragments were measured directly, while $Mo_{1-x}W_xS_2$ required exfoliation to obtain homogeneous regions. The alloy compositions were determined based on the positions of the (00.6) and (00.2) reflections for $Mo(S_{1-x}Se_x)_2$ and $Mo_{1-x}W_xS_2$, respectively, assuming a linear variation of lattice parameters between the binary endpoints (i.e., $MoS_2$–$MoSe_2$ and $MoS_2$–$WS_2$).